\documentclass{PoS}

\usepackage{epsf,amssymb,amscd,amsmath}

\def\TriangleOS{{\color{red}\;\raisebox{-20mm}{\epsfysize=57mm\epsfbox{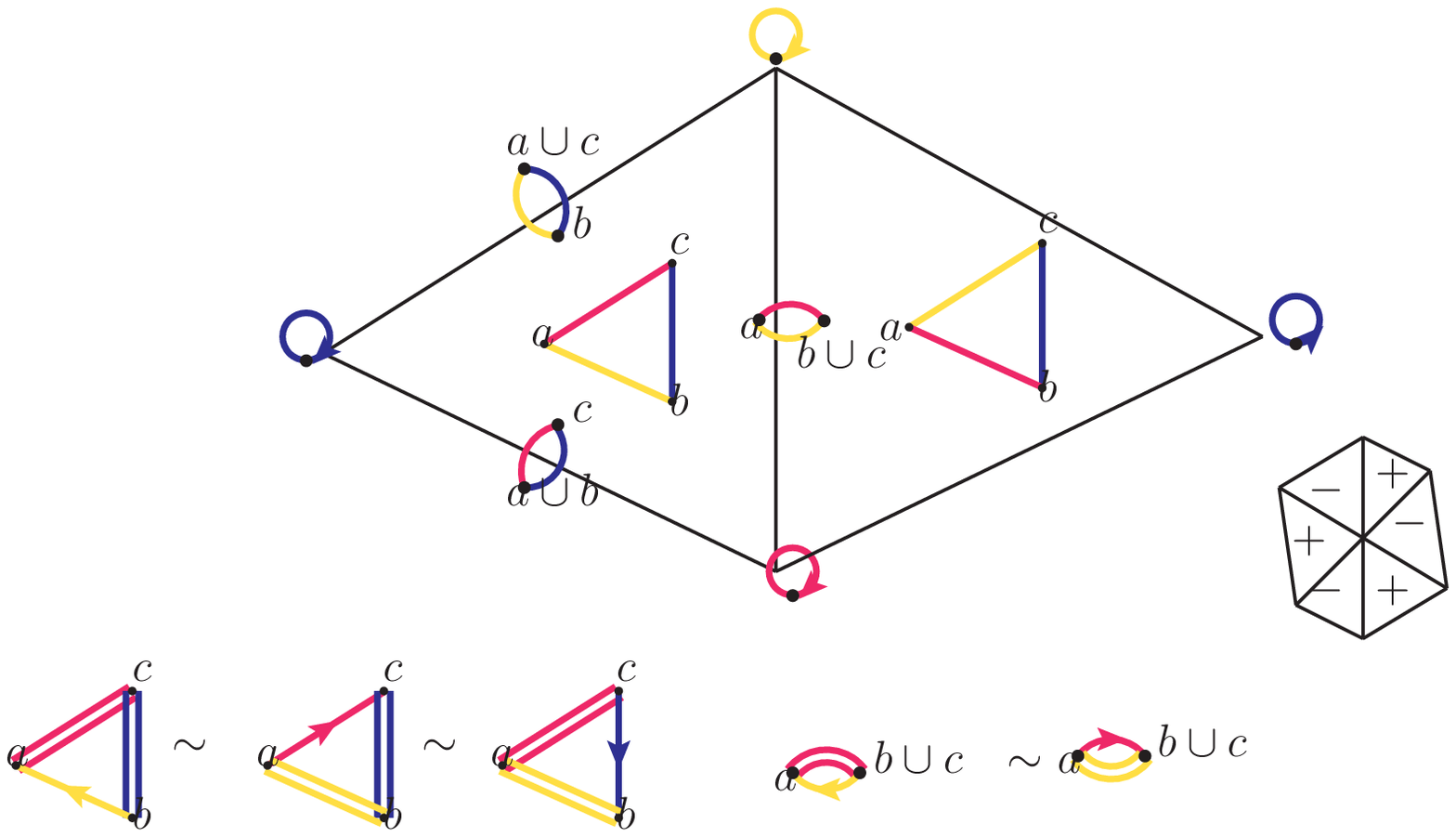}}\;}}
\def\corollaDiff{{\color{red}\;\raisebox{-20mm}{\epsfysize=40mm\epsfbox{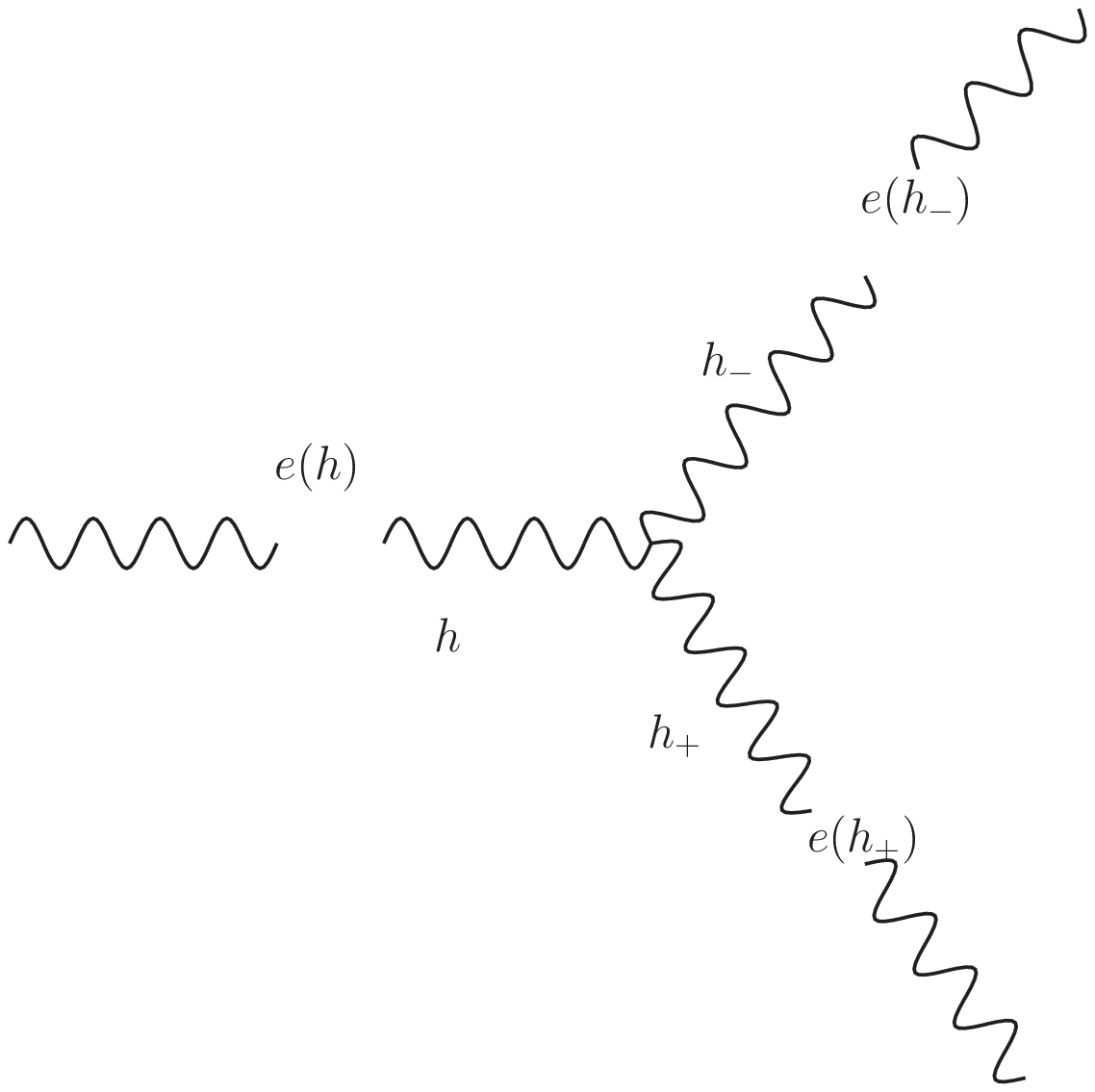}}\;}}
\def\triangleQCD{{\color{red}\;\raisebox{-20mm}{\epsfysize=60mm\epsfbox{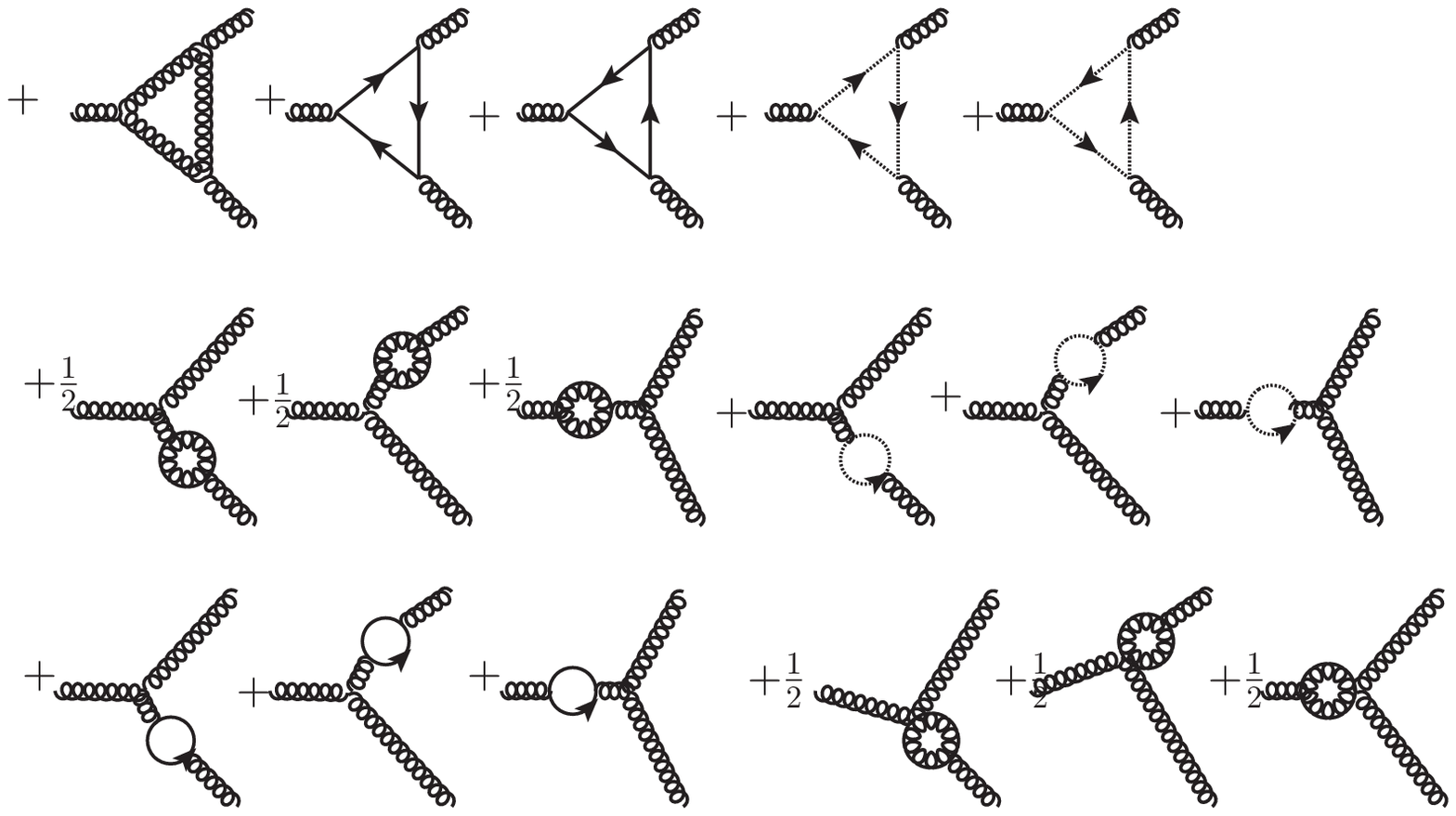}}\;}}
\def\triangleScalar{{\color{red}\;\raisebox{-20mm}{\epsfysize=30mm\epsfbox{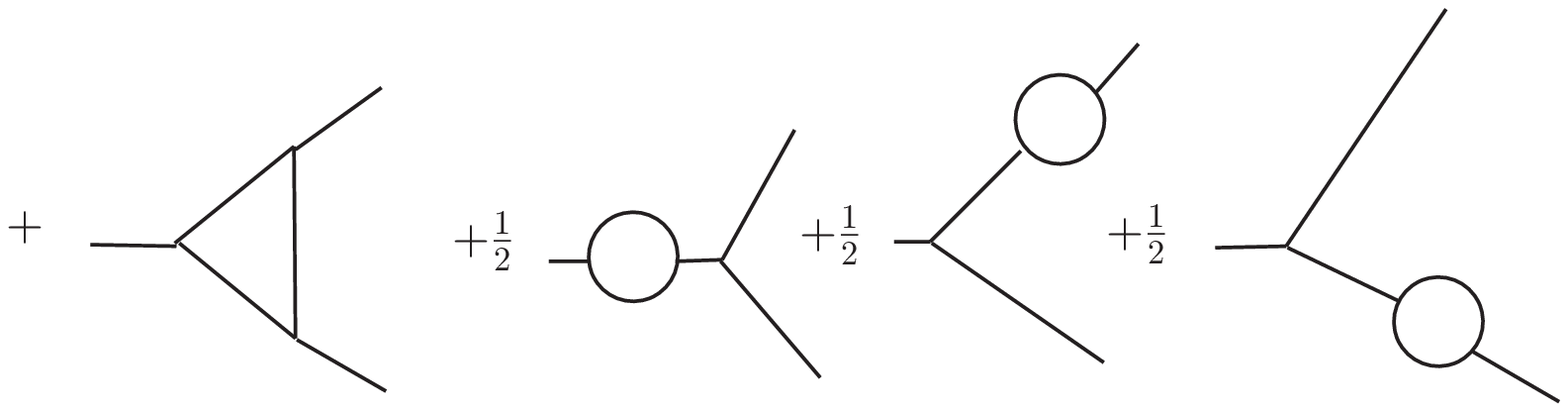}}\;}}

\title{The corolla polynomial: a graph polynomial on half-edges}

\ShortTitle{The corolla polynomial}

\author{\speaker{Dirk Kreimer}\thanks{Thanks to Johannes Bl\"umlein and Peter Marquard for organizing our workshop.}\\
        Depts.\ of Physics and Mathematics\\ Humboldt University, Berlin\\
        E-mail: \email{kreimer@physik.hu-berlin.de}}

\abstract{The study of Feynman rules is much facilitated by the two Symanzik polynomials, homogeneous polynomials based on edge variables for a given Feynman graph. We review here the role of a recently discovered third graph polynomial based on half-edges which facilitates the transition from scalar to gauge theory amplitudes: the corolla polynomial. We review in particular the use of graph homology in the construction of this polynomial.}

\FullConference{Loops and Legs in Quantum Field Theory (LL2018)\\
		29 April 2018 - 04 May 2018\\
		St. Goar, Germany}

\begin{document}

\section{Introduction}
The computation of a gauge theory suffers the enormous number of integrals one is confronted with. This results from the fact that the number of Feynman graphs contributing to a given amplitude is much bigger than in a scalar field theory, and from the fact that the spin structure of the Feynman rules gives a much more complicated tensorial structure.

Hence, the reduction to master integrals is a much more cumbersome undertaking. Here we report on recent progress connecting the integrand for a connected amplitude in non-abelian  gauge theory to the amplitude for a scalar connected 3-regular amplitude: all vertices are cubic.

\section{Scalar vs Gauge Theory amplitudes}
The starting point is the amplitude for a Feynman graph in a scalar cubic theory which is given (in $D=4$ dimensions say) through the two graph- or Symanzik-polynomials $\Phi_\Gamma,\psi_\Gamma$, where 
\[
\Phi_\Gamma=\phi_\Gamma+\left(\sum_{e\in E_\Gamma} A_em_e^2\right)\psi_\Gamma,
\] 
and $\psi_\Gamma,\phi_\Gamma$ are the classic first and second graph polynomials:
\[
\psi_\Gamma=\sum_{\mathrm{spanning\,trees}\, T } \prod_{e\not\in T} A_e,
\]
\[
\phi_\Gamma=\sum_{\mathrm{spanning\,two-trees}\, T_1\cup T_2 } Q(T_1)\cdot Q(T_2)\prod_{e\not\in T_1\cup T_2} A_e,
\]
with $Q(T_i)$ the sum of the external momenta attached to vertices of $T_i$.
 
There exists then  a corolla differential $D_\Gamma$ such that the Feynman integrand $I_\Gamma$ for connected 3-regular graphs $\Gamma$,
\[
I_\Gamma=\frac{e^{-\frac{\Phi_\Gamma}{\psi_\Gamma}}}{\psi_\Gamma^2},\label{scalint}
\]
gives rise, when summed over connected 3-regular  graphs $\Gamma$
to the total gauge theory amplitude, using $D_\Gamma I_\Gamma$.

For an example consider the connected one-loop triangle graphs.
The scalar graphs are, 
$$\triangleScalar$$
whilst in the gauge theory case we have more:
$$\triangleQCD$$
We have internal quark- and ghost-loops, and 4-valent gluon vertices.

If we replace first all edges in the 3-regular  scalar case by gauge-boson edges there are two steps remaining: to either shrink internal boson edges to generate 4-valent boson vcertices, or else to replace internal gauge-boson loops by ghost or fermion loops.

It is remarkable that in this process, the ranks of the automorphism groups of graphs play along (meaning that symmetry factors play along). This is due to an underlying double complex of two graph homologies based on either shrinking edges or marking closed cycles in the graph \cite{KSvS}. 

Both homologies can be implemented using a new graph polynomial on half-edges, the corolla polynomial \cite{KY}.
\section{Graph Homology}
\noindent Theorem (\cite{KSvS}):\\
Let $e$ be an edge connecting two 3-gluon  vertices in a graph $\Gamma$, $\chi^e_+$ be the operator which shrinks edge $e$, extend $\chi_+^e$ to zero when acting between any other two vertices.
Let $S$, with $S^2=0$,  be the corresponding graph homology operator. Then, for a gauge theory amplitude $r$:\\
Let $X^{r,n}_{0\mathrm{x};j \mathrm{gl}}$ be the sum of all 3-regular connected graphs, with $j$ ghost loops,   and with external legs determined by $r$
and loop number $n$, weighted by colour and symmetry, let $X^{r,n}_{/\mathrm{x},j\mathrm{gl}}$
be the same allowing for 3- and 4-valent vertices.
We have\\
\[
i):\;e^{\chi_+}X^{r,n}_{0\ \mathrm{x};j\mathrm{gl}} = X^{r,n}_{/\mathrm{x};j\mathrm{gl}},\]
\[
ii):\;Se^{\chi_+}X^{r,n}_{0\mathrm{x};j\mathrm{gl}}=0.
\]\\[7mm]
This theorem shows we can generate all 4-valent couplings by graph homology as studied in Vogtmann's paper \cite{CV} based on a study of graph homology initiated by Kontsevich.

\noindent Theorem (\cite{KSvS}):\\
Let $\delta^C_+$ be the operator which marks a cycle
$C$ through 3-valent vertices and unmarked edges, extend $\delta_+^C$ to zero on any other cycle.
Let $T$, with $T^2=0$,  be the corresponding cycle homology operator. Then:

Let $X^{r,n}_{j\mathrm{x};0\mathrm{gl}}$ be the sum of all connected graphs with $j$ 4-vertices contributing to amplitude $r$ and loop number $n$ and no ghost loops, weighted by colour and symmetry, $X^{r,n}_{j\mathrm{x};/\mathrm{gl} }$
be the same allowing for any possible number of ghost loops.
We have\\
\[
i):\; e^{\delta_+}X^{r,n}_{j\mathrm{x};0\mathrm{gl}}=X^{r,n}_{j\mathrm{x};/\mathrm{gl}},\]
\[
ii):\; Te^{\delta_+}X^{r,n}_{j\mathrm{x};0\mathrm{gl}}=0.\]\\[7mm]
This theorem ensures that we cn insert ghost loops (or fermion loops for that matter) using a cycle homology on graphs. This homology was introduced in \cite{KSvS}.

These two operations are compatible \cite{KSvS,Knispel}:

\noindent Theorem (\cite{KSvS}):\\
\begin{itemize}
\item[i)] We have $[S,T]=0\Leftrightarrow (S+T)^2=0$ and 
$$
Te^{\delta_++\chi_+}X^{r,n}_{0\mathrm{x};0\mathrm{gl}}=0, \qquad Se^{\delta_++\chi_+}X^{r,n}_{0\mathrm{x};0\mathrm{gl}}=0.
$$
\item[ii)] Together, they generate the whole gauge theory amplitude from 3-regular graphs:
\[ e^{\delta_++\chi_+}X^{r,n}_{0\mathrm{x};0\mathrm{gl}}=X^{r,n}_{/\mathrm{x};/\mathrm{gl}}=:X^{r,n}.
\]
\end{itemize}

$X^{r,n}$, the series over all graphs contributing to a physical amplitude, each graph weighted by its symmetry factor,  is the only non-trivial element in the bicomplex of cycle- and graph-homology \cite{KSvS,Knispel}. This indeed is BRST homology graph-theoretically. To interprete BRST homology by graph homologies is an approach very much in line with an early analysis of QCD provided for example in Predrag Cvitanovic's lecture notes \cite{Cvitanovic}. 
\section{The corolla polynomial and differential}
The next step is to use the above structures to find an efficient transition from scalar to gauge theory amplitudes. This is facilitated by the corolla polynomial given in \cite{KY} whose definition we follow:

It is a polynomial based on half-edge variables
$a_{v,j}$ assigned to any half-edge $(v,j)$ determined by a vertex $v$ and an edge $j$
of a graph $\Gamma$.
We have
\begin{itemize}
\item For a vertex $v\in V$ let $n(v)$ be the set of edges incident to $v$ (internal or external).
\item For a vertex $v\in V$ let $D_v = \sum_{j \in n(v)} a_{v,j}$.
\item Let $\mathcal{C}$ be the set of all cycles of $\Gamma$ (cycles, not circuits). This is a finite set.
\item For $C$ a cycle and $v$ a vertex in $V$, since $\Gamma$ is 3-regular, there is a unique edge of $\Gamma$ incident to $v$ and not in $C$, let $v_C$ be this edge.
\item For $i\geq 0$ let 
  \[
  C^i = \sum_{\substack{C_1,C_2,\ldots C_i \in \mathcal{C} \\ C_j \text{pairwise disjoint}}} \left(\left(\prod_{j=1}^{i} \prod_{v \in C_j}a_{v,v_C}\right)\prod_{v \not\in C_1\cup C_2\cup \cdots \cup C_i} D_v\right)
  \]
\item Then
  \[
   C = \sum_{j \geq 0} (-1)^j C^j
  \]
\end{itemize}
$C=C_\Gamma$ is the corolla polynomial. 

Having this polynomial at our disposal, we can replace any of its half-edges by a differential operator. This defines a corolla differential which acts on the scalar integrand.  It acts on auxiliary external momenta $\xi_e$ (which are set to the physical momenta only after the action) which we supplement in the scalar integrand for each internal edge $e$.
\begin{equation*} D_{g}(h) := -\tfrac12 g^{\mu_{h_+}\mu_{h_-}} \Big( \epsilon_{h_+}\frac1{A_{e(h_+)}}\frac\partial{\partial\xi(h_+)_{\mu_h}}-\epsilon_{h_-}\frac1{A_{e(h_-)}}\frac\partial{\partial\xi(h_-)_{\mu_h}} \Big), \end{equation*}
for any half-edge $h$. Here, a half-edge $h$ determines at a 3-regular vertex two other half-edges $h_+,h_-$ using that we can orient each corolla by a theorem in \cite{CV}
much used in \cite{KSvS} (which in tun determines the signs $\epsilon_{h_\pm}$), and each such half-edge furthermore determines an edge $e=e(h)$ to which it belongs. 
$$\corollaDiff$$
Double differentials wrt the same half edge generate the Feynman rules for a 4-valent vertex via Cauchy's residue formula: differentiating twice, the Leibniz rule ensures the emergence of poles with residues which are the contributions of grapgs with 4-valent gluon vertices \cite{KSvS}.
\section{Results}
Finally, we get the Feynman integrands in the unrenormalized and renormalized case for a gauge theory amplitude $r$ from 3-regular connected graphs of scalar fields.

\noindent Theorem (\cite{KSvS}):\\
The full Yang--Mills amplitude $\bar U_\Gamma$ for a graph $\Gamma$ can be obtained by acting with a corolla differential operator on the scalar integrand $U_\Gamma (\{\xi_e\})$ for $\Gamma$, setting the edge momenta $\xi_e = 0$ afterwards.\\  
Moreover, $\bar U_\Gamma$ gives rise to a differential form $J_\Gamma^{\bar U_\Gamma}$ and there exists a vector $H_\Gamma$ such that the unrenormalized Feynman integrand for the sum of all Feynman graphs contributing to the connected $k$-loop amplitude $r$ is  
\[
\Phi(X^{r,k})=
\sum_{|\Gamma|=k,\mathbf{res}(\Gamma)=r} \frac{\mathrm{colour}(\gamma)}{\mathrm{sym}(\Gamma)}
\int_{H_\Gamma} J_\Gamma^{\bar{U}_\Gamma},\] 
The renormalized analogue is given by writing $\bar{U}_\Gamma^R$ instead of $\bar{U}_\Gamma$ \cite{KSvS}.\\[7mm]
This was all generalized to the full Standard Model by David Prinz \cite{Prinz}, showing that even spontaneous symmetry breaking respects the structure of graph homologies and can be approached by the corolla polynomial, with suitable adoptions for the matter and ghost content of that model.

\section{Outer Space Structure of Gauge Theory}
There are growing signs that the parametric approach to Feynman diagrams has a deep connection to the structure of Outer Space \cite{BKOS,KOS}, in particular when investigating monodromies of amplitudes. For the case of gauge theories where vertices of valence higher than four vanish, this generates rather interesting Outer Spaces where cells corresponding to co-dimension two hypersurfaces are missing.

A simple example is the one-loop triangle graph. The co-dimension one edges are boundaries populated by a graph with one three-valent and one four-valent vertex (unions like $a\cup b$ in the figure). The zero-dimensional co-dimension two cells are points decorated by tadpoles which would have a forbidden 5-valent vertex $a\cup b\cup c$.
This is to be removed.
$$\TriangleOS$$

Note that when such Outer Spaces with missing cells adopted to gauge theories have been constructed, 
a Green Function is an integral over the whole such Outer Space - a piecewise linear sum of integrals \cite{Berghoff} over the volume of all cells of the still allowed  codimensions.
\section{Outlook}
\begin{itemize}
\item The two Kirchhoff polynomials are distinguished as unique polynomials on edge variables having recusive contraction deletion properties.
\item The corolla polynomial is similarly distinguished amongst half-edge polynomials
having recursive half-edge deletion properties \cite{KY}.
\item This allows to construct the renormalized integrand for a connected amplitude in gauge theory from the scalar amplitudes for connected 3-regular graphs \cite{KSvS}.
\item David Prinz has generalized this approach to amplitudes to the full SM \cite{Prinz}.
\item Marcel Golz is turning this into a very efficient algorithm for QED amplitudes \cite{Golz}.
\item What is the corolla polynomial for spin 2 bosons
and hence for quantum gavity?
\item Can we combine the reduction to master integrals (Laporta's algorithm) with this transition to gauge theory?
\end{itemize}

\end{document}